\documentclass[a4paper,fleqn,usenatbib]{mnras}
\usepackage{graphicx,amssymb,amsmath,times,verbatim}
\usepackage{color}
\usepackage{subfigure}
\usepackage{upgreek}

\usepackage{tabularx,ragged2e}
\usepackage{longtable}
\usepackage{supertabular}

\usepackage{adjustbox}
\usepackage{lipsum}
\usepackage{pdflscape}

\title[Log-normal flux distributions]{On the determination of log-normal flux
distributions for astrophysical systems}

\author[Z. Shah et al.]{Zahir Shah$^{1}$\thanks{Email: zahir@iucaa.in},  Ranjeev Misra$^{1}$ and Atreyee Sinha$^2$ \\
$^{1}$ Inter-University Center for Astronomy and Astrophysics, PB No.4, Ganeshkhind, Pune-411007, India\\
$^2$  Laboratoire Univers et Particules de Montpellier, Universit\'e Montpellier, CNRS/IN2P3, CC 72, Place Eug\'ene Bataillon, F-34095 Montpellier Cedex 5, France}

\begin{document}
\date{}
\pagerange{\pageref{firstpage}--\pageref{lastpage}} \pubyear{2020}
\maketitle
\label{firstpage}
\begin{abstract}
Determining whether the flux distribution of an  Astrophysical
  source is a Gaussian or a log-normal, provides key insight into
  the nature of its variability. For lightcurves of moderate length
  ($< 10^3$), a useful first analysis is to test the Gaussianity of
  the flux and logarithm of the flux, by estimating the skewness and
  applying the Anderson-Darling (AD) method. We perform extensive simulations
  of lightcurves with different lengths, variability, Gaussian measurement
  errors and power spectrum index $\beta$ (i.e. $P(f) \propto f^{-\beta}$), to provide a prescription
  and guidelines for reliable use of these two tests. We present empirical
  fits for the expected standard deviation of skewness and
  tabulated AD test critical
  values for $\beta = 0.5$ and $1.0$, which differ from the values given in
  the literature which are for white noise ($\beta = 0$). Moreover,
  we show that for white noise, for most practical situations,
  these tests are meaningless, since binning in time alters
  the flux distribution. For  $\beta \gtrsim 1.5$, the skewness variance does not decrease with length and hence the tests are not reliable. Thus, such tests can
  be applied only to systems with $\beta\gtrsim 0.5$ and  $\beta \lesssim 1.0$.
  As an example of  the prescription given in this work, we reconfirm
  that the Fermi data of the blazar, 3FGL\,J0730.2-1141,  shows that
  its $\gamma$ ray flux is consistent
  with a log-normal distribution and not with a Gaussian one.

\end{abstract}

\section{Introduction}

Recent advancements in observational astronomy have now made  long-term
continuous observation of astrophysical sources possible. These
observations are crucial for obtaining consistent and conclusive
interpretation of flux distributions.  The study of long-term flux
distributions is an important tool for analysing and characterising
the stochastic variations in  astrophysical sources.
Characterisation of flux distribution can provide insight into the
 underlying physical process which drives the
variability in the source. For example, the central limit theorem
states that the addition of a larger number of independent random variables will
result in a Gaussian distribution. Therefore, a Gaussian distribution of
flux would imply additive models, where a linear summation
of components contributes to the emission. On the other hand, a log-normal
distribution (which is a Gaussian distribution when the flux is taken in log scale) would suggest that the
mechanism driving the emission is of a multiplicative nature \citep{Lyubarskii1997, Uttley2005}.

Over the past decade,  log-normal flux distributions have been
observed  in  X-ray light curves of compact accreting system like
X-ray binaries and Active Galactic Nuclei (AGN)
\citep{Uttley2001,Vaughan2003,Gaskell2004,Uttley2005}. The emission in
these sources is mainly produced by accretion discs.  This possibly connects
the log-normal behaviour of observed flux distributions to the
accretion disc. 
The log-normal flux variations in the  accretion disc have been
explained by the propagation fluctuation model
\citep{Lyubarskii1997}. In this model,  fluctuations in the mass
accretion rate are produced on timescale corresponding to the local viscous time
scales. They propagate inward and couple together
to produce a multiplicative behaviour in the inner part of disk
\citep{Lyubarskii1997,Uttley2001,Uttley2005,Arevalo2006,McHardy2010}. 
Log-normal behaviour of flux distribution has also been observed in
sources which are dominated by the non-thermal jet emission
\citep{Giebels2009,Ackermann2015,Sinha2016, Shah2018, Meyer2019, Khatoon2020}. One
possible realisation for such observations is that  disk fluctuations
are possibly imprinted on the jet emission
\citep{Giebels2009,McHardy2010,Shah2018} i.e. the fluctuations from the disk
are efficiently transmitted to the jet and thereby modulating the jet
emission accordingly.
However, the minute time scale variation observed in high energy light
curves \citep{Gaidos1996,Aharonian2007,Paliya2015} 
reflect that the jet emission should be independent of accretion disc
fluctuations \citep{Narayan2012}. In such cases, the log-normal
distribution in flux can be produced by the linear Gaussian
perturbations in the intrinsic particle acceleration time-scales in
the acceleration region \citep{Sinha2018}. Moreover, the fluctuations
in the diffuse escape time scales of the emitting electrons would
produce flux distribution shapes other than Gaussian or
log-normal. Alternatively, \citet{Biteau2012} have shown that under
specific conditions, the additive shot noise-model can also produce a
log-normal flux distribution. For example the Doppler boosting of 
emission from a large number of randomly oriented mini-jets 
results in a flux distribution with features similar to that of a 
log-normal distribution.

Two of the common tests used to differentiate between a Gaussian and
a log-normal distribution are the skewness \citep{Zwillinger2000} and the Anderson-Darling (AD) \citep{Stephens1974}.
These tests can be used to determine at some confidence level (a) that the
distribution is not consistent with a Gaussian one and (b) that it
is consistent with a log-normal one. Thus, they are first step prelude
for more detailed investigations such as direct fitting of the binned
distributions. Also, they can be implemented on data with a relatively
small number of points. However, the reliability and effectiveness of
the tests depend on several aspects of the flux light curve such as
the amplitude of the variability, the length of the data and the percentage
of measurement error that the data has. Further, the slope of the power spectral density which characterizes the variability 
also has an important impact on the reliability of these tests. In this work, using simulated
data, we study these effects on the effectiveness and reliability of the
tests. Our aim is to quantitatively understand these effects and provide
guidelines to a user as to the minimum number of points (as a function
of the variability strength and measurement errors)  required
for an effective implementation of AD and skewness test. The parameter values corresponding to different
confidence levels for the  AD test, have been numerically
computed in the literature for the case when the lightcurve has an underlying
power spectrum independent of frequency i.e. for white noise. However, lightcurves
often have power spectrum with a power-law frequency dependence and we tabulate
the corresponding parameter values for different confidence levels as
a function of the power-law index. Thus we facilitate the correct use of the
test to realistic lightcurves. Moreover, there are subtle and important
issues regarding the effect of time-binning of the data on the nature 
of the resultant flux distribution. In general the binning changes nature of non-Gaussian distribution such as log-normal one, i.e. an intrinsic log-normal distribution may not remain a log-normal when the light curve is binned. Since the observed lightcurves are always binned to some time-bin, it is not clear why several astrophysical systems display log-normal distributions? In this work, we address this paradoxical issue by analysing the simulated lightcurves with the skewness and AD test.

The framework of this paper is as following:-   In section (\S\ref{sec:skew_AD_white}), we carry a detailed investigation of the lightcurves derived from white noise, for the effective and reliable use of skewness and AD test to check the Gaussianity/log-normality of flux distribution. In section (\S\ref{sec:PL}), we explore the efficiency of the skewness and AD tests for light curves with power-law power spectra. We applied the derived results of our simulation to observed flux distribution of one of bright  \emph{Fermi} blazar  in section \S\ref{sec:sim_app}.  In section \S\ref{sec:conc}, we provide summary and discussion of the work with important guidelines for the effective use of skewness and AD test.

\section{Systems with white noise}\label{sec:skew_AD_white}

Skewness is measure of the asymmetry of a distribution and its value  can be zero, positive or negative. For a flux distribution of $N$ points, the sample skewness, which is third standardised moment is computed as the Fisher-Pearson coefficient i.e., 
\begin{equation}
\kappa_F=\frac{\frac{1}{N}\sum_{i\rightarrow 1}^{N}(x_i-\bar{x})^3}{s^3}
\end{equation}
where $\bar{x}$ and $s^2$ are mean and variance of the sample, these two quantities are first and second statistical moments, respectively.
In case of symmetric distribution
  such as a Gaussian one, the skewness value will be zero. Thus, if the
  skewness of a flux distribution is shown to be inconsistent with zero,
  while the skewness of its log flux distribution is shown to be
  consistent with zero, then one can ascertain that the underlying
  distribution is consistent with a log-normal one and inconsistent
  with a Gaussian one.\\
  In addition to skewness test, Anderson Darling (AD) test is also used to check the Gaussianity or log-normality of the flux distribution. AD test is a statistical test which checks the null hypothesis that a sample is drawn from a particular distribution (e.g, normal distribution). The AD test returns the test statistic ($TS$-value) and critical value ($cv$). The $cv$  are function of the number of points in the lightcurve and are obtained numerically at significance levels of 15\%, 10\%, 5\%, 2.5\%, 1\%. The null hypothesis that the sample is drawn from normal distribution is rejected at a significance level, if the returned TS values is greater than the critical value at that significance level. We have used $cv$ at 5\% significance level in this work, which implies that we accept the null hypothesis at 95\% confidence level.  In this section, we consider the case when the distribution is derived from white noise i.e. when the power spectrum is independent of frequency.

\begin{figure*}
\centering
\includegraphics[width=0.48\textwidth,angle=270]{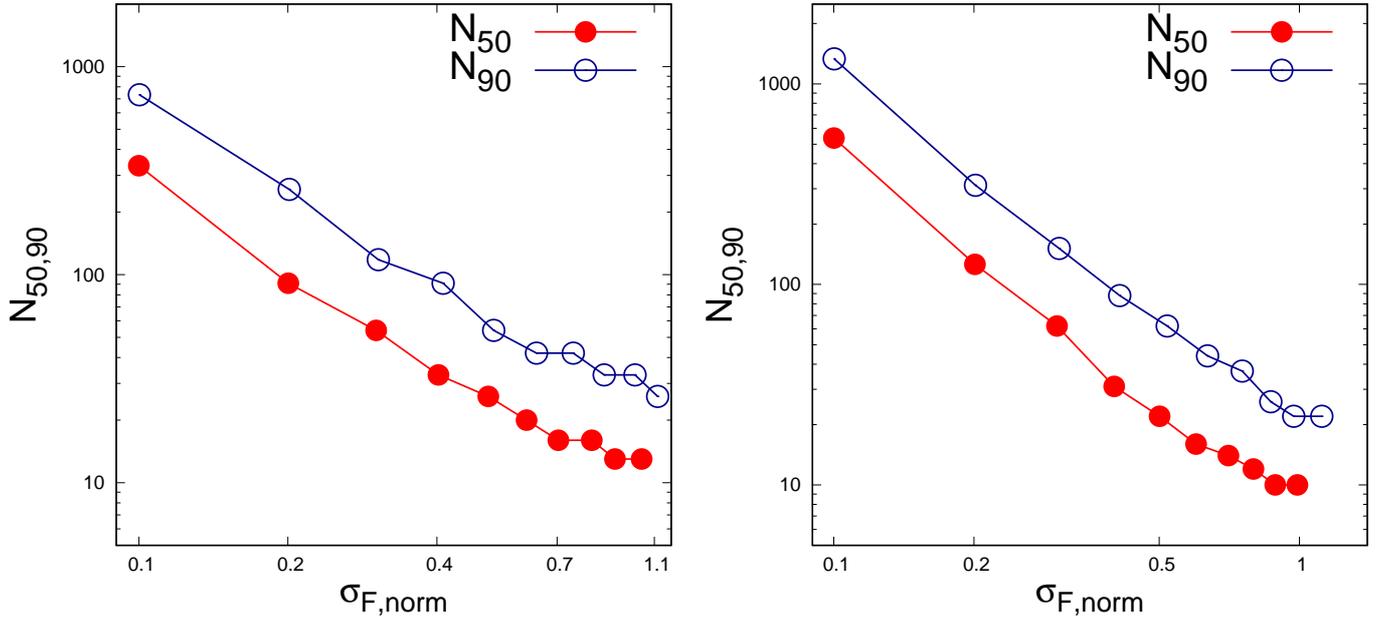}

\caption{Left Panel: The minimum number of bins required,  $N_{50}$ and $N_{90}$,
  for 50\% and 90\%  of the lightcurves to have $\Delta\kappa_{F} > 0$
  as a function of normalised standard deviation $\sigma_{F,norm}$. Right Panel: Same as the
  left one except that the condition is for the AD test, i.e. $TS_{diff} > 0$. 
}
\label{fig:skew_ad_number}
\end{figure*}

  For a finite length lightcurve from a Gaussian flux distribution,
  the measured skewness, $\kappa_F$  will deviate
  from zero with a standard deviation of $\sqrt(6/n)$ \citep{Tabachnick1996}, where $n$ is
  the number of bins in the lightcurve.  A user would then
  check if  the skewness difference of a given lightcurve
\begin{equation}\label{eq:skew_diff}
\Delta\kappa_F=|\kappa_F|-L\sqrt{\frac{6}{n}}
\end{equation}
exceeds zero for
a certain confidence, $L$ then its true  $\kappa_F$ is nonzero. Since it is sensible to use 95\% confidence level, therefore we carried the analysis for $L=2$ (i.e. 2-sigma confidence). For a
lightcurve whose intrinsic distribution is log-normal, $\Delta\kappa_F$
would exceed zero only for a sufficiently large number of points, $n$.
To quantify the minimum number of points required, we simulated
lightcurves corresponding to a log-normal distribution.
In practice, we generated a time-series of the logarithm flux from
  a Gaussian distribution  with centroid $\mu_{LF} = 0$ and some values
  of width  $\sigma_{LF}$ and length $n$, and then converted it to a
  flux lightcurve by taking the exponential of each value. Given the values of $\mu_{LF}$ and $\sigma_{LF}$, one can obtain the theoretical estimate of mean of log-normal distribution $\mu_F$, using $\mu_{F}=\exp(\mu_{LF}+\frac{\sigma_{LF}}{2})$.
  For different sets of  $\sigma_{LF}$ and $n$, we simulated 3000 such
  flux lightcurves, where $\sigma_{LF}$ ranges from 0.1--1. For each of the 3000 simulated
  lightcurves corresponding to a fixed value of  $\sigma_{LF}$ and length $n$, we
  computed the mean $\mu_{F}$, square root of the variance
  $\sigma_{F}$  and the skewness difference $\Delta\kappa_{F}$. The fraction,
  $f$ of the 3000 lightcurves for which  $\Delta\kappa_{F} > 0$ was noted,
  which increased as the number of bins in the lightcurve, $n$ increased.
  Figure \ref{fig:skew_ad_number} shows the minimum number of bins required,  $N_{50}$ and $N_{90}$,
  for 50\% and 90\%  of the lightcurves to have $\Delta\kappa_{F} > 0$,
  as a function of the average normalised standard deviation, defined as
  \begin{equation}\label{eq:sigma_norm}
\sigma_{F,norm}=\frac{\bar{\sigma_F}}{\bar{\mu_F}}
  \end{equation}
 
  The Figure \ref{fig:skew_ad_number} should be interpreted as 
  giving an estimate of the
  minimum number of bins a lightcurve should have in order to
  use skewness to reject the hypothesis that its flux distribution is
  Gaussian. The minimum number of bins required for 90\%  of the lightcurves to have $\Delta\kappa_{F} > 0$ ranges from 25 for $\sigma_{F,norm} \sim 1.1$ to 730 for $\sigma_{F,norm} \sim 0.1$.
  
  While,  $\Delta\kappa_{F}$ can be used to reject the hypothesis that the
  lightcurve is from a Gaussian distribution, in principle one can go further
  and compare the measured skewness $\kappa_{F}$ with the theoretical value
  for a log-normal distribution,
\begin{equation}\label{eq:pred_skew}
\kappa_T=\frac{\sigma_F^3}{\mu_F^3}+\frac{3\sigma_F}{\mu_F}
\end{equation}
However, from our simulation we found that this is not reliable. The
measured value of skewness deviates significantly from the theoretical value
for  finite number of
data points and large $\sigma_F$. The comparison between the
observed and theoretical skewness values are given in appendix
\ref{apx:SS}.

\begin{figure*}
\centering
\subfigure{\includegraphics[scale=0.49,angle=270]{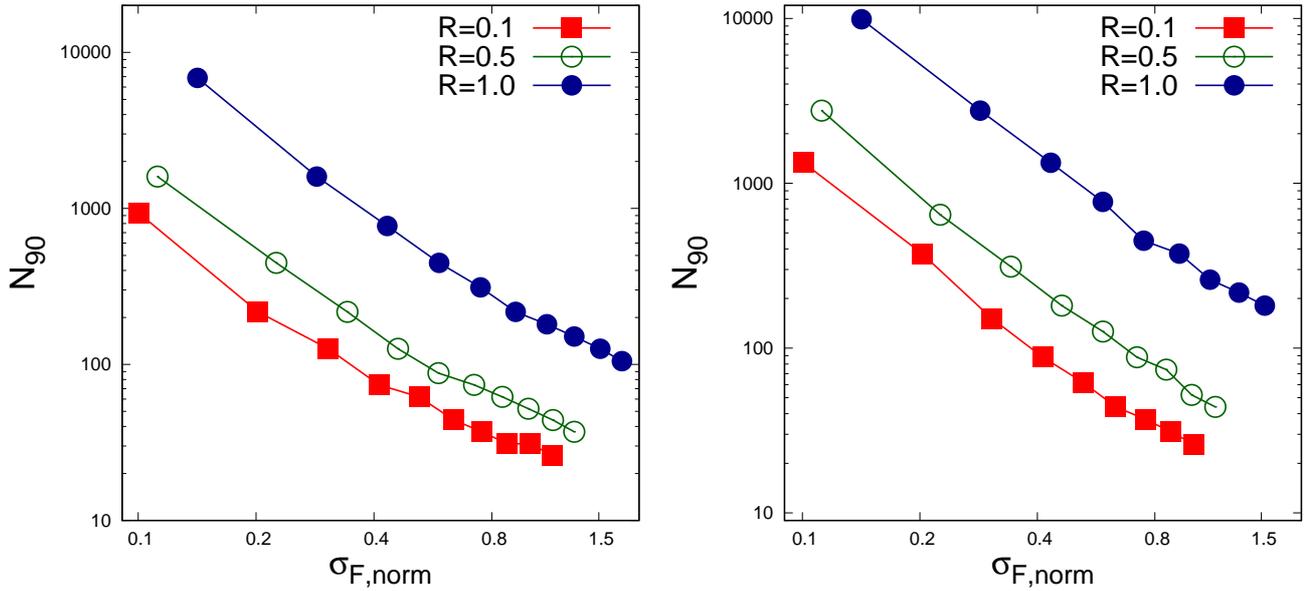}}
%\subfigure{\includegraphics[scale=0.32,angle=270]{sigma_skew_0.5_2SE.eps}}\\
\caption{Left panel: Minimum number of flux points, $N_{90}$, required for 90\% of light curves to have $\Delta \kappa_F>0$, is plotted as function of  normalize standard deviation  $\sigma_{F,norm}$. The three curves with points shown by colours red (filled square), green (open circle) and blue (filled circle) are for error fraction, R=0.1, 0.5 and 1 respectively. Right Panel: Same as left one except the condition is for the AD test i.e. $TS_{diff}>0$.}
\label{fig:skew_ad_frac_error}
\end{figure*}

Measurement errors of the flux can have a significant impact on the skewness
test. If the errors are significant, it would
require a larger number of flux points for a distribution obtained from an intrinsically log-normal
one, to reject the hypothesis that the distribution is Gaussian. To quantify
this effect, we introduce measurement errors which are normally distributed
to the simulated lightcurves. We  define a parameter, error ratio, $R$ as the
ratio of the sigma of the error distribution $\sigma_{err}$ to that of
the intrinsic flux $\sigma_{F}$.   The measurement error is generated as random normal distribution with standard deviation, $\sigma_{err}=R*\sigma_F$. We have added this random normal distribution to the intrinsic simulated log-normal light curves. The AD and skewness tests are then applied to the modified log-normal distribution (combination of random noise and intrinsic log-normal distribution). Figure \ref{fig:skew_ad_frac_error} shows
the variation of $N_{90}$ (left panel)  as a function
of average normalised standard deviation $\sigma_{F.norm}$ for $R = 0.1, 0.5$
and $1.0$. The Figures shows that if the error fraction is larger than 0.5,
the number of points in the lightcurve required to reject the Gaussianity of
the distribution is significantly larger than the case when there are no
measurement errors.

\begin{figure*}
%\centering
\subfigure{\includegraphics[scale=0.48,angle=270]{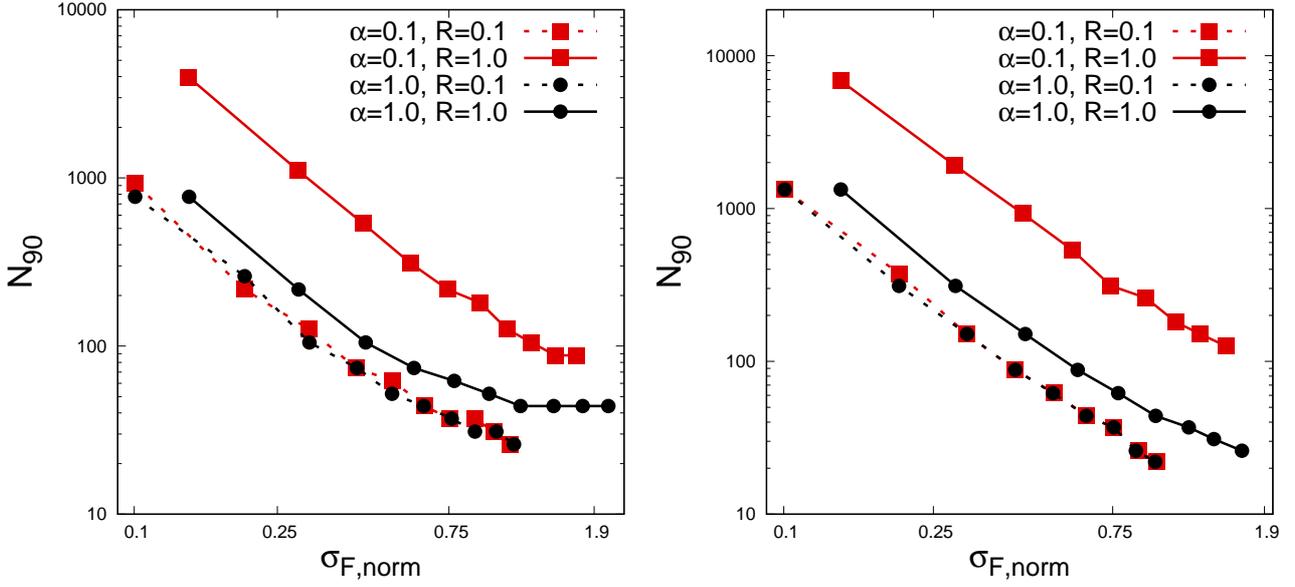}}\quad \hspace{-0.5cm}
\caption{Minimum number of flux bins, $N_{90}$,  as function of normalised standard deviation of flux
  distribution $\sigma_{F,norm}$,  when the measurement error depend on the flux value.  The left and right panel are obtained using skewness and AD tests respectively. In both panels, black and red curves correspond to $\alpha=$ 0.1 and 1.0 respectively. The dashed and solid representation of curves are for R= 0.1 and 1.0 respectively.}
\label{fig:skew_sigma_number_error_slope}
\end{figure*}

Moreover, in some circumstances the measurement error on the flux depends on the
flux value, $F_i$ i.e. $\sigma_{err}\propto {F_i}^{\alpha}$. We have simulated
lightcurves for such cases and show the result for $\alpha = 0.1$ and $1.0$
in Figure \ref{fig:skew_sigma_number_error_slope}. Here $N_{90}$ is plotted
against $\sigma_{F.norm}$ for $R = 0.1$ and $1.0$. For $\alpha = 1.0$, the
number of points required to reject the Gaussianity is less than for $\alpha = 0$.

Similarly,  we repeated the above analysis for the AD test. For a lightcurve, we define a measure
\begin{equation}\label{eq:ts_diff}
TS_{diff.}=TS-cv_{5\%}
\end{equation}
where $cv_{5\%}$ is the critical value such that the probability that TS is
larger for a Gaussian distribution is 5\%. Analogous to the
skewness test, we use simulations to estimate $N_{90}$ ($N_{50}$),
the minimum number of bins for which $90\%$ ($50\%$) of the simulated
lightcurves have a $TS_{diff} > 0$. The results are presented in the right
panels of Figures \ref{fig:skew_ad_number},  \ref{fig:skew_ad_frac_error} and \ref{fig:skew_sigma_number_error_slope}.

For a log-normal distribution, the skewness and AD test will accept
the normality of the logarithm of the flux at more than 95\% confidence level by definition. However, if there is
measurement error in the flux, then this may not be so and for sufficiently
large error and large number of data points, the test would indicate that
the log of the flux distribution is not Gaussian. We simulated lightcurves
with measurement errors and find that when the number of points exceeds
$10^{5}$ ($10^4$), the skewness test fails when the error ratio is 0.2 (0.5).
Similar results are obtained for the AD test. 

As a thumb rule, our results show that both  skewness and AD tests can be
used to determine that the
distribution is not Gaussian, for lightcurves having more than 100 points and with
a normalised standard deviation (fractional r.m.s of variability) greater
than 30\%. The ratio of the sigma of the measurement error to that of
the intrinsic variability should be less $0.2$. The Figures presented can
be used to gauge the reliability of the tests for other values. 

\begin{figure}
%\centering
\subfigure{\includegraphics[scale=0.33,angle=270]{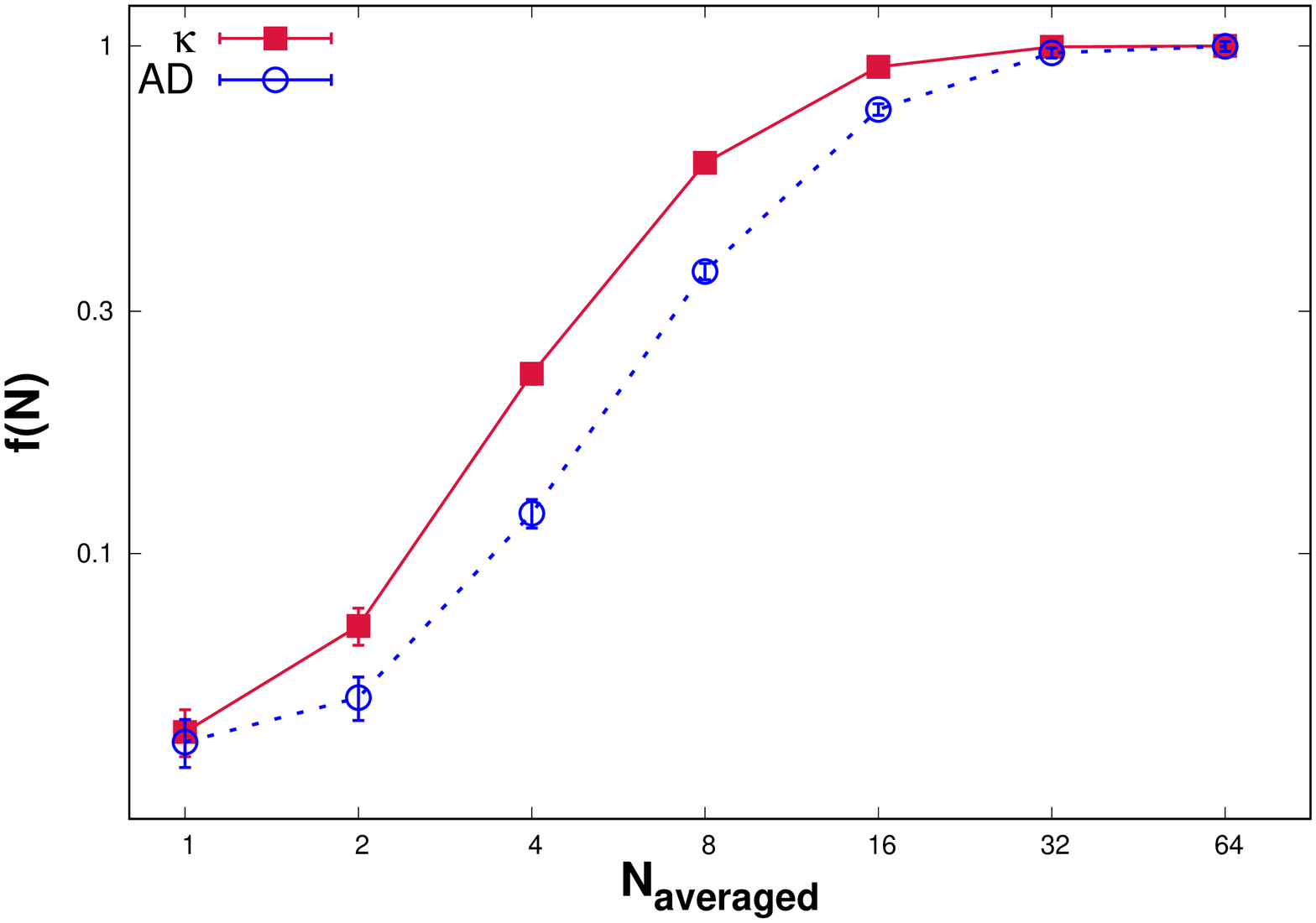}}\quad \hspace{-0.5cm}
\caption{Top panel: The fraction for which log-normality is rejected in the binned flux distribution $f(N)$ as function of the number of flux points over which averaging is done. %in order to obtain light curve with 400 flux points and $\sigma_L=0.3$.  
The solid red curve with filled square points corresponds to skewness test, while the dotted blue curve with open circles corresponds to AD test.}
\label{fig:flux_bin_skew_ad}
\end{figure}

The measurement errors for a lightcurve can be decreased by binning the
lightcurve in time. Moreover, most measurement processes involves summing
the fluxes for a minimum time which is the time resolution of the instrument.
While a Gaussian distribution remains a Gaussian one with binning in time, this
is not the case for an intrinsically log-normal one. We have performed simulations to illustrate this point quantitatively. We binned  simulated lightcurves from a log-normal distributions such that the resultant binned lightcurve has
400 data points and a $\sigma_L \sim 0.3$. The logarithm of the fluxes of these
binned lightcurves were then subjected to the skewness and AD test, and the
fraction of the lightcurves which failed the tests were noted. This fraction
for the skewness and AD test are plotted as a function of the number points
averaged in Figure \ref{fig:flux_bin_skew_ad}. It is clear from the Figure that binning changes the
intrinsic log distribution to some other distribution.

Thus, an intrinsically ``true'' log-normal distribution which has a white noise
power spectrum cannot be determined to be such, if the lightcurve has been binned. If there is a cutoff in the power spectrum such that there is no intrinsic
variability on time-scales shorter than a particular value and if the time
resolution is smaller than this value, then the distribution maybe determined to
be log-normal. This is further discussed in the last section.

\section{Systems with power-law noise}\label{sec:PL}

For many physical systems, the power spectra generated from the lightcurves
can be described as a power-law i.e. $P(f) \propto f^{-\beta}$. In this section,
we explore the efficiency of the skewness and AD tests for such
systems. Random time series representing power-law
power spectra and whose distribution are Gaussian can be generated
using the algorithm proposed by \citet{Timmer1995}.  Using this algorithm, we generated a large sample of light curves for different sets of $n$ and $\beta$, the minimum and maximum frequency of the corresponding power law power spectra are chosen as $f_{min}=\frac{1}{n}$ and $f_{max}=0.5$ , respectively. For $\beta\leq1$, the time-series generated  with \citet{Timmer1995} algorithm will have Gaussian PDF for all values of $n$  \citep{Morris2019} . However, for steeper power spectral index ($\beta>1$),
the light curves becomes more non-stationary and generated time-series are less likely to be described by Gaussian PDF \citep{Alston2019, Morris2019}. Therefore, in this section we mainly focus on the results for $\beta\lesssim1$.

In the previous section, the analysis was based on systems with white
noise i.e. for the special case of $\beta = 0$. It should be noted that
 the standard deviation of
skewness as well as the critical values for the AD tests
are valid only for white noise  and cannot be used 
for the general  non-zero $\beta$ cases. Since we did not find these
values in the Literature, we first undertook a simulation study to
estimate them.

\begin{figure}
\centering
\subfigure{\includegraphics[scale=0.32,angle=270]{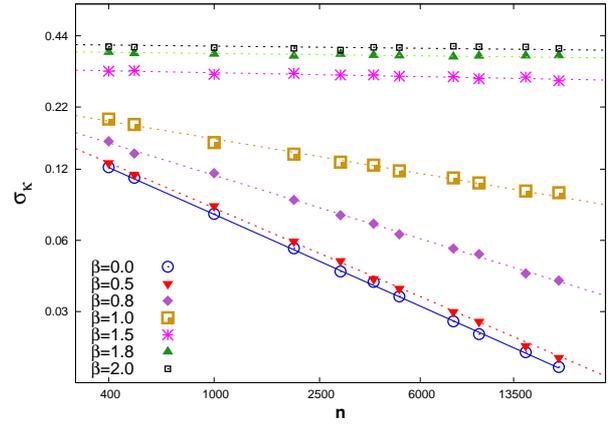}}\quad \hspace{-0.5cm}
\caption{The standard deviation of skewness if plotted against the number of flux points for different values of the power spectrum
index $\beta$. The solid line is the expected standard deviation $\sigma_\kappa = \sqrt{6/n}$ valid for the case when $\beta = 0$.
The dotted lines are the fitted lines using Equation \ref{eq:skew_err}, with best fit values tabulated in Table \ref{table:skew_err}.}
\label{fig:std_skew_err}
\end{figure}

In order to estimate standard error of skewness for nonzero $\beta$-cases, we simulated lightcurves having power law PSD  for different sets of $n$ and $\beta$. For each pair of $n$ and $\beta$, we generated 2000 random light curves and measured the skewness value in each trial. We plotted the standard deviation of skewness, $\sigma_k$ (standard deviation obtained from 2000 skewness
values) with $n$ in Figure \ref{fig:std_skew_err}.  For $\beta = 0$, the variation in $\sigma_k$ with
$n$ is as expected i.e.  $\sigma_\kappa = \sqrt(6/n)$ and this is shown as
a solid line. For $\beta < 0.5$ the standard deviation of skewness is close to that
for $\beta = 0$, but for larger $\beta$ there is significant deviation.
We empirically fit the variation with $n$ using a generic function
\begin{equation}\label{eq:skew_err}
\sigma_\kappa (n)=(a/n)^b
\end{equation}
and tabulate the best fit values of $a$ and $b$ for different values
of $\beta$ in Table \ref{table:skew_err}.
The fits are shown by dotted lines in Figure \ref{fig:std_skew_err}.
For a stationary time-series, the error (i.e. the standard deviation
  of the skewness) should decrease with the number of elements, such that for
  a long enough series , the skewness is well defined and convergent.
  This is indeed confirmed here for white noise ($\beta = 0$) which is expected to be
  stationary.   For $\beta\lesssim1$, the slope of standard deviation of skewness with $n$ gets flattened (i.e. its dependence on $n$ decreases) showing that
  the system is approaching non-stationarity.
Further, for $\beta>1$, \citet{Morris2019} have shown that the light curves generated with the \citet{Timmer1995} algorithm are less likely to be Gaussian due non-stationarity. This is further established here as in such cases the standard deviation in skewness remains almost constant with $n$. For example, as shown in Figure \ref{fig:std_skew_err}, for $\beta\gtrsim 1.5$, $\sigma_{\kappa}$ is independent of $n$. Thus for such cases the tests are not reliable. The  divergence of tests for $\beta>1$ can thus be mainly attributed to increase in non-stationarity, where the variance of individual lightcurve segments is not constant in time but instead varies about a mean value determined by the underlying PSD \citep{Uttley2005}.

\begin{table}
%\begin{center}\begin{tiny}
\centering
%\large
\begin{tabular}{lcr}
\hline 
% $\beta$ &  a  &  b &  $\kappa_{err.}(n)$ \\
$\beta$ &  a  &  b \\
\hline
\hline
0.0 &  5.96$\pm$0.17 & 0.499$\pm$0.003\\
0.1 &  5.63$\pm$0.21 & 0.494$\pm$0.003\\
0.2 &  6.86$\pm$0.25 & 0.513$\pm$0.004\\
0.3 &  5.89$\pm$0.27 & 0.497$\pm$0.004\\
0.4 &  6.04$\pm$0.33 & 0.496$\pm$0.005\\
0.5 &  5.58$\pm$0.24 & 0.482$\pm$0.004\\
0.6 &  4.54$\pm$0.23 & 0.451$\pm$0.004\\
0.7 &  3.92$\pm$0.19 & 0.418$\pm$0.004\\
0.8 &  1.81$\pm$0.18 & 0.345$\pm$0.005\\
0.9 &  0.58$\pm$0.07 & 0.269$\pm$0.004\\
1.0 &  0.06$\pm$0.02 & 0.189$\pm$0.005\\

\hline
\hline
\end{tabular}
\caption{Skewness error fit parameters `a' and `b' obtained by fitting the equation (\ref{eq:skew_err}) to the variation of skewness standard deviation with $n$ for different values of $\beta$.}
\label{table:skew_err}
\end{table}

Similarly using 2000 simulated lightcurves,  we estimated the critical
values for the AD test for different values of $\beta$ and
significance levels as a function of the number of bins and the results
are tabulated in Table \ref{table:critical_val}. As a check we note that the critical values obtained
for $\beta=0$ are similar to those given in Literature (written in parenthesis
in Table  \ref{table:critical_val}).

\begin{figure*}
%\centering
\subfigure{\includegraphics[scale=0.48,angle=270]{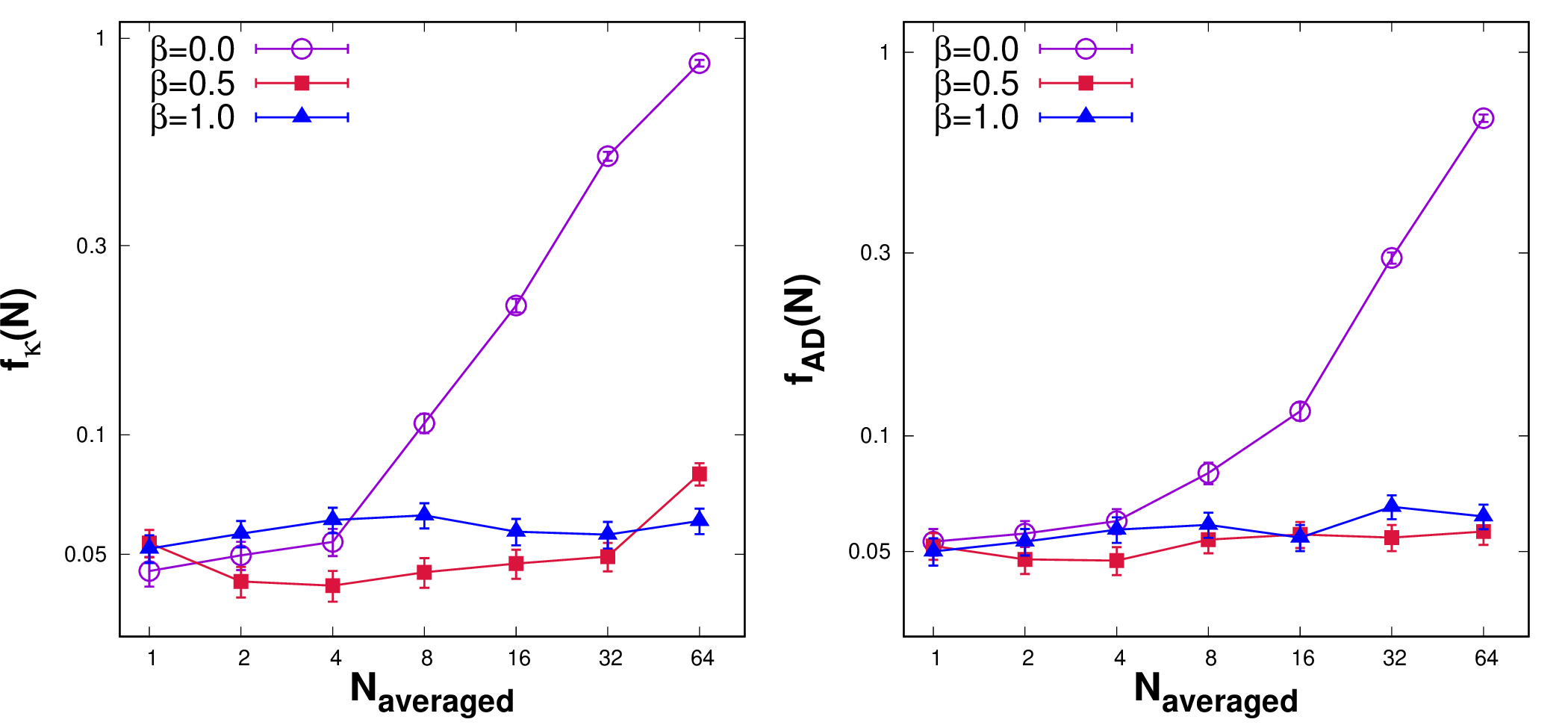}}
\caption{Left panel:The fraction $f_{\kappa}(N)$ for which log-normality of the binned light curves is rejected ($\Delta \kappa>0$), plotted against the  number of averaged bins. The length of each binned light curve is 400.  
The three curves viz. violet curve with open circle points, red curve with filled square points and blue curve with filled triangle points corresponds to the light curves which are generated from  power law noise with $\beta=0.0$, 0.5 and 1.0 respectively. Right panel: Same as the left one except the condition used is for AD test.}
\label{fig:flux_pl_bin_skew_ad}
\end{figure*}

Having obtained the standard deviation for skewness and the critical values
of AD test applicable for lightcurves having power-law power spectra, we
perform similar analysis as was done in the previous section for white noise.
In the end of the previous section it was shown that if a lightcurve
having a log-normal distribution and white noise, is binned, the resultant
distribution is no longer a log-normal one and both the skewness and AD test
will reject the hypotheses that it is log-normal. We first test the effect of
binning on lightcurves generated from a power-law distribution. We simulated a sample of Gaussian time-series using \citet{Timmer1995} algorithm, which were exponentiated in order to obtain lightcurves having log-normal distribution. The exponential model reproduces the observed properties for both stationary and non-stationary light curves \citep{Alstonet2019, Alston2019}. It should be noted that the shape of power spectra may not be preserved by exponentiating a time-series. However, \citet{Uttley2005} showed that the effect is negligible for PSD's observed in astrophysical system and it is acceptable to consider the shape of power spectra of log-normal distribution same as that of input linear time-series. The generated log-normal lightcurves are binned in order to have lightcurves with 400 points and a $\sigma_L \sim 0.3$. The distribution of the logarithm of the fluxes were tested using
skewness and AD tests. In Figure \ref{fig:flux_pl_bin_skew_ad}, the fraction of the lightcurves which failed the tests are plotted as a function of the binning factor. As expected
for no binning ($N_{average} = 1$), the fraction $f_{\kappa}$ and $f_{AD}$ is 0.05
since a 95\% confidence has been taken for the tests. As shown earlier,
for white noise ($\beta = 0$), the fractions increase with binning showing that
the distribution is no longer consistent with being a log-normal. However,
for $\beta = 0.5$ and $\beta = 1.0$, the fractions are more or less independent
of binning indicating that the log-normal nature is maintained after binning
for systems where the power spectrum has a power-law shape with index greater than 0.5. A plausible reason that binning preserves the log-normality could be related to low power in high frequency components for lightcurve having steeper PSD.  The maximum frequency in case of binned lightcurve can be written as $f_{max}=\frac{1}{2N_{averaged}\Delta t}$, where $\Delta t$ is the time bin of unbinned (or original) light curve, the maximum frequency decreases as the binning (or $N_{averaged}$) increases.  In case of steeper PSD, most power is contained in low frequency components and less power will be at higher frequencies. In such cases, the effect of truncation of high frequencies will be less, hence the properties of the original light curve (e.g, log-normal) are preserved. However, in case of white noise, the variability at higher frequencies strongly influence the behaviour of the time-series, hence removal of high frequencies from the PSD will change the time series property effectively.

\begin{figure*}
\centering
\includegraphics[width=0.46\textwidth,angle=270]{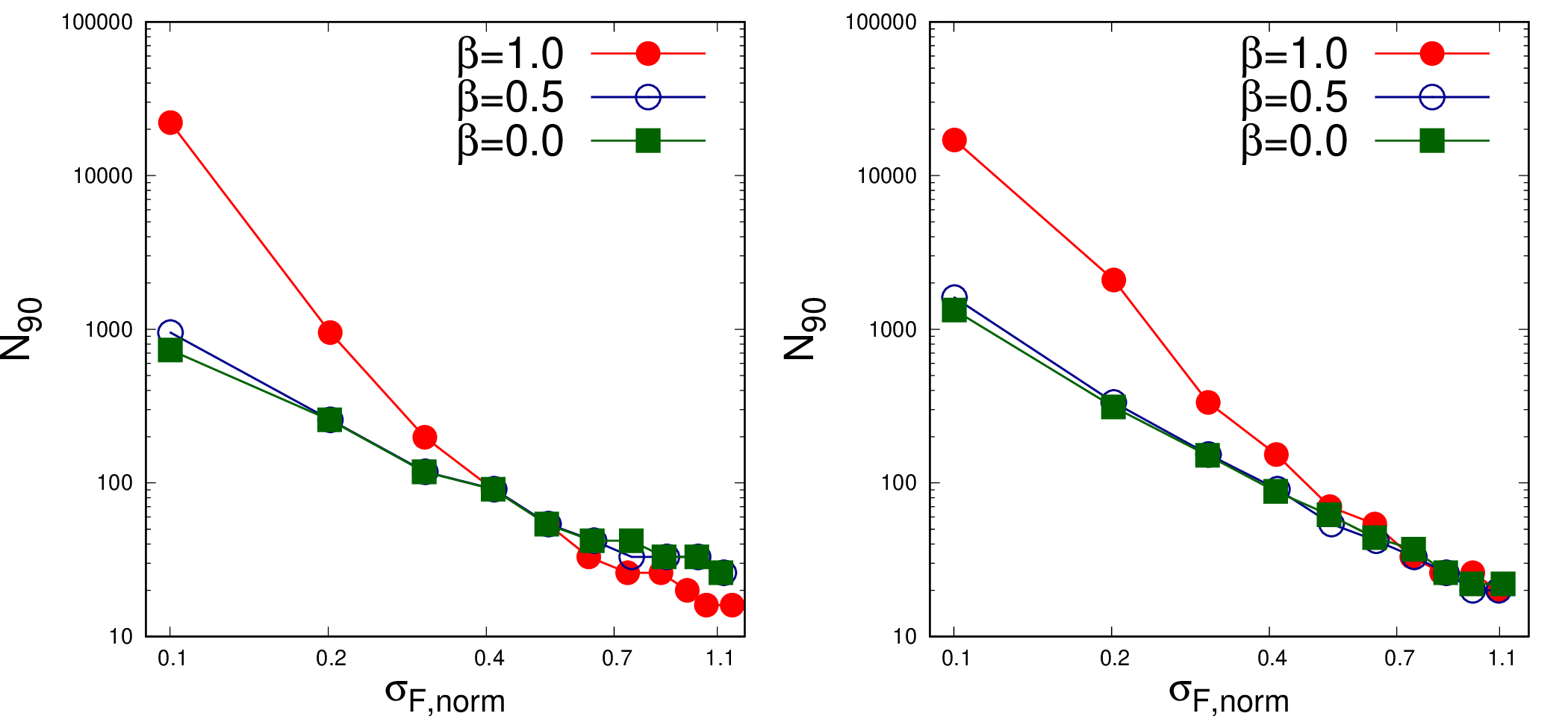}

\caption{The minimum number of flux points required,  $N_{90}$ such that
  90\% of the lightcurve generated will  have $\Delta\kappa_{F} > 0$ (left panel) and $TS_{diff}>0$ (right panel)
  as a function of normalised $\sigma_{F,norm}$. In both panels, the three curves viz. green curve with filled squares, blue curve with open circles and red curve with filled circles corresponds to $\beta$=0.0, 0.5 and 1.0 respectively.   
}
\label{fig:skew_frac}
\end{figure*}

\begin{figure*}
\centering
\subfigure{\includegraphics[scale=0.495,angle=270]{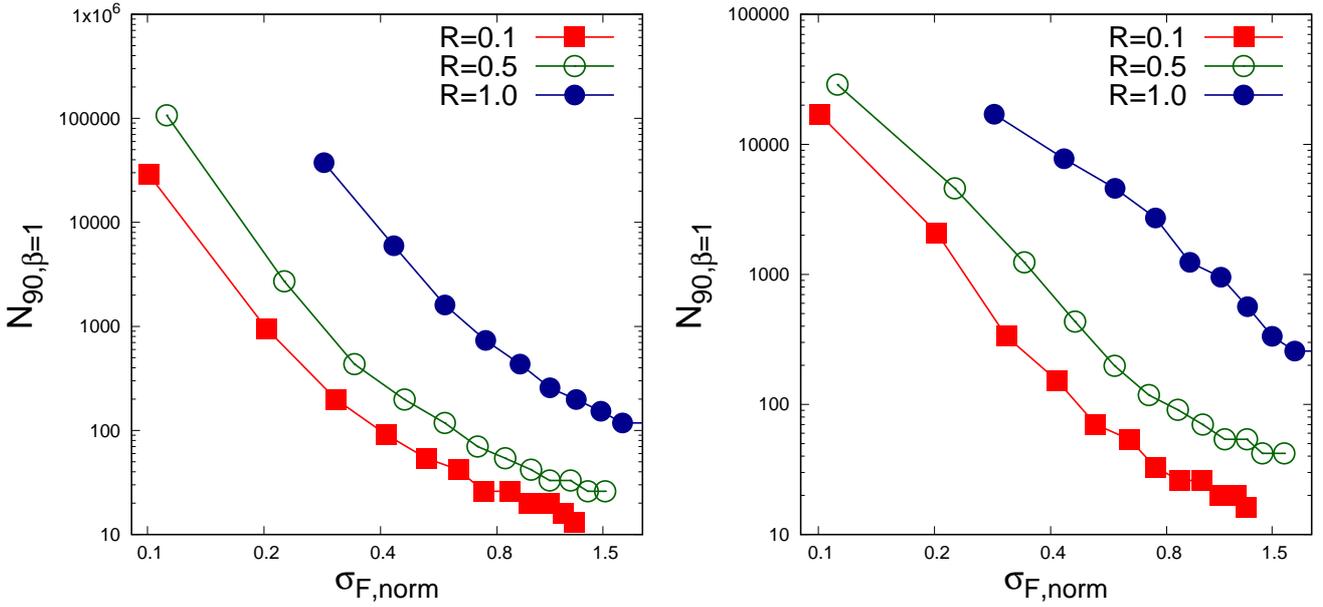}}
%\subfigure{\includegraphics[scale=0.32,angle=270]{sigma_skew_0.5_2SE.eps}}\\
\caption{Left panel: Minimum number of flux points, $N_{90}$, required for 90\% of the light curves generated from power law noise with $\beta=1.0$, to have $\Delta \kappa_F>0$, is plotted as function of  normalise standard deviation  $\sigma_{F,norm}$. The three curve's with points shown by colours red (filled square), green (open circle) and blue (filled circle) are for error fraction, R=0.1, 0.5 and 1 respectively. Right Panel: Same as left one except the condition  $TS_{diff}>0$ used is for the case AD test.}
\label{fig:skew_frac_error}
\end{figure*}

We continue the analysis by estimating the minimum number of bins required to
reject the hypothesis that the flux distribution of such lightcurves are Gaussian.
The left panel of Figure \ref{fig:skew_frac}, displays the minimum number of points required for a log-normal lightcurve, such that for 90\% of them $\Delta \kappa_F > 0$, $N_{90}$, as a
function of the normalised standard deviation.
The plots are for $\beta = 1.0$, $0.5$ and $0.0$, where $\beta = 0.0$ is
a repetition of the results obtained in the previous section for white noise.
The right panel of Figure \ref{fig:skew_frac}, is the same as the left except that it is
for the case when $TS_{diff} >0$ for the AD test. We note that the results are
similar for $\beta = 0.5$ and white noise, but there is marked difference for
$\beta = 1.0$, where a substantially larger number of bins is required to
reject Gaussianity for normalised standard deviations less than $0.5$ as
compared to white noise. The effect of measurement error is shown in
Figure \ref{fig:skew_frac_error}, where $N_{90}$ as a function of the normalised standard deviation
is shown for different values of the error ratio, $R$ for $\beta = 1.0$. 
As expected increasing the ratio of measurement errors to the intrinsic
variation, $R = \sigma_{err}/\sigma_F$), requires a larger number of data points
to reject Gaussianity.

Thus, the skewness and AD tests to identify a system has a log-normal
distribution can only be effectively used if the power spectra  has
an index, $\beta \gtrsim 0.5$, otherwise flux binning would modify the
distribution. On the other hand, for large $\beta \sim  1$, the number
of bins required to reject the hypothesis that the distribution is Gaussian,
is significantly larger ($\gtrsim1000$), than for lower values
of $\beta$. The most effective analysis can be undertaken when
$\beta \sim 0.5$, error ration $R<0.2$, the normalised standard deviation of the lightcurve
is $\gtrsim 30\%$ and the number of bins is $> 100$.

%\begin{landscape}
\begin{table*}
\caption{critical values of AD test for power law noise with  $\beta=0.0, 0.5, 1.0$ obtained from simulations. The critical values are obtained at significance level  of 15\%, 10\%, 5\%, 2.5\% and 1\% .}
\label{table:critical_val}
%\begin{center}\begin{tiny}
%\centering
%\large
\begin{adjustbox}{width=0.4\textwidth,center=\textwidth}
\begin{tabular}{lccc}
\hline 
 Slope & number of & significance  & CV calculated  (CV given)  \\
($\beta$) &	 data points (n)	&  level	\%	& 	\\	
\hline
\hline
%0.0 & 8192 & 2000	&	15	&	0.59	&	 0.21	&	0.57\\
0.0 & 4096   & 	15	&	0.56 (0.575)	\\
0.0 & 2048	&	15	&	0.57 (0.575)	\\
0.0 & 1024 	&	15	&	0.56 (0.574)	\\
0.0 & 512  	&       15	&	0.56 (0.572) \\
0.0 & 100	&	15	&	0.55 (0.555) \\
\hline
%0.0 & 8192 & 2000	&	10	&	0.66	&	 0.21	&	0.656\\
0.0 & 4096 	&	10	&	0.62 (0.655)	\\
0.0 & 2048   &	10	&	0.63 (0.655)\\
0.0 & 1024 	&	10	&	0.63 (0.653)\\
0.0 & 512 	&       10	&	0.63 (0.651)\\
0.0 & 100	&	10	&	0.68 (0.632)\\
\hline
%0.0 & 8192 & 2000	&	5	&	0.80	&	 0.21	&	0.787\\
0.0  & 4096 	&	5	&	0.75 (0.786)	\\
0.0  & 2048 	&	5	&	0.74 (0.785)\\
0.0  & 1024 	&	5	&	0.75 (0.784)\\
0.0  & 512  	&       5	&	0.75 (0.781)\\
0.0 & 100 & 5 & 0.75 (0.759) \\

\hline
%0.0 & 8192 & 2000	&	2.5	&	0.92	&	 0.21	&	0.918\\
0.0  & 4096 	&	2.5	&	0.88 (0.917)	\\
0.0  & 2048 	&	2.5	&	0.87 (0.916)\\
0.0  & 1024	&	2.5	&	0.89 (0.914)\\
0.0  & 512  	&       2.5	&	0.87 (0.911)	\\
0.0  & 100  	&       2.5	&	0.86 (0.885) \\
\hline
%0.0 & 8192 & 2000	&	2.5	&	0.92	&	 0.21	&	0.918\\
0.0  & 4096 &	1.0	&	1.02 (1.091)	\\
0.0  & 2048 	&	1.0	&	1.04 (1.09)	\\
0.0  & 1024 &	1.0	&	1.06 (1.088)\\
0.0  & 512  	&       1.0	&	1.03 (1.084)\\
0.0  & 100  	&       1.0	&	1.02 (1.053)\\
\hline 
\hline
%0.5 & 8192 & 2000	&	15	&	0.59	&	 0.21	&	0.57\\
0.5 & 4096  &	15	&	0.58\\
0.5 & 2048  &	15	&	0.58	\\
0.5 & 1024  &	15	&	0.58	\\
0.5 & 512  &  15	&	0.57	\\
0.5 & 100	&	15	&	0.56 \\
\hline
%0.5 & 8192 & 2000	&	10	&	0.66	&	 0.21	&	0.656\\
0.5 & 4096 &  10	&	0.66	\\
0.5 & 2048 &  10	&	0.65	\\
0.5 & 1024 &  10	&	0.65	\\
0.5 & 512  &   10	&	0.65	\\
0.5 & 100  &   10	&	0.68 \\
\hline
%0.5 & 8192 & 2000	&	5	&	0.80	&	 0.21	&	0.787\\
0.5 & 4096 & 5	&	0.80	\\
0.5 & 2048 & 	5	&	0.79	\\
0.5 & 1024 &	5	&	0.79	\\
0.5 & 512  &  5	&	0.78	\\
0.5 & 100 & 5 & 0.76 \\
\hline
%0.5 & 8192 & 2000	&	2.5	&	0.92	&	 0.21	&	0.918\\
0.5 & 4096 & 2.5	&	0.94	\\
0.5 & 2048 & 	2.5	&	0.93	\\
0.5 & 1024 & 	2.5	&	0.92	\\
0.5 & 512  &   2.5	&	0.91	\\
0.5 & 100  &   2.5	&	0.88 \\
\hline
%0.5 & 8192 & 2000	&	2.5	&	0.92	&	 0.21	&	0.918\\
0.5 & 4096 & 	1.0	&	1.11	\\
0.5 & 2048 & 	1.0	&	1.12	\\
0.5 & 1024 &	1.0	&	1.09	\\
0.5 & 512  &  1.0	&	1.04	\\
0.5 & 100  &  1.0	&	1.05 \\
\hline
\hline
%0.5 & 8192 & 2000	&	15	&	0.59	&	 0.21	&	0.57\\
1.0 & 4096 & 	15	&	2.62	\\
1.0 & 2048 & 	15	&	2.09	\\
1.0 & 1024 & 	15	&	1.57	\\
1.0 & 512  &       15	&	1.15	\\
1.0	& 100	&	15	&	0.69 \\
\hline
%0.5 & 8192 & 2000	&	10	&	0.66	&	 0.21	&	0.656\\
1.0 & 4096 & 	10	&	3.30	\\
1.0 & 2048 & 	10	&	2.65	\\
1.0 & 1024 & 	10	&	1.95	\\
1.0 & 512  &       10	&	1.40	\\
1.0 & 100  &       10	&	0.67 \\
\hline
%0.5 & 8192 & 2000	&	5	&	0.80	&	 0.21	&	0.787\\
1.0 & 4096 & 	5	&	4.70	\\
1.0 & 2048 & 	5	&	3.72	\\
1.0 & 1024 &   5	&	2.68	\\
1.0 & 512  &        5	&	1.64	\\
1.0 & 100 & 5 & 0.98 \\
\hline
%0.5 & 8192 & 2000	&	2.5	&	0.92	&	 0.21	&	0.918\\
1.0 & 4096 & 	2.5	&	5.85	\\
1.0 & 2048 & 	2.5	&	4.83	\\
1.0 & 1024 & 	2.5	&	3.47	\\
1.0 & 512  &       2.5	&	2.42	\\
1.0 & 100  &       2.5	&	1.17	\\
\hline
%0.5 & 8192 & 2000	&	2.5	&	0.92	&	 0.21	&	0.918\\
1.0 & 4096 & 	1.0	&	7.76	\\
1.0 & 2048 &    1.0	&	6.67	\\
1.0 & 1024 & 	1.0	&	4.90	\\
1.0 & 512  &       1.0	&	3.26	\\
1.0 & 100  &       1.0	&	1.43   \\
\hline
\hline
\\
\end{tabular}
\end{adjustbox}
\end{table*}
%\end{landscape}

% Ranjeev pointed out that AD critical values and KS probability values are valid only for power law noise with exponent ($\beta=0$). So we are calculating the new critical and p-values for $\beta=0.5, 1.0$.\\

\section {Application of simulation results on blazar $\gamma$-ray observations}\label{sec:sim_app}
\begin{figure}
\centering
\includegraphics[width=0.34\textwidth,angle=270]{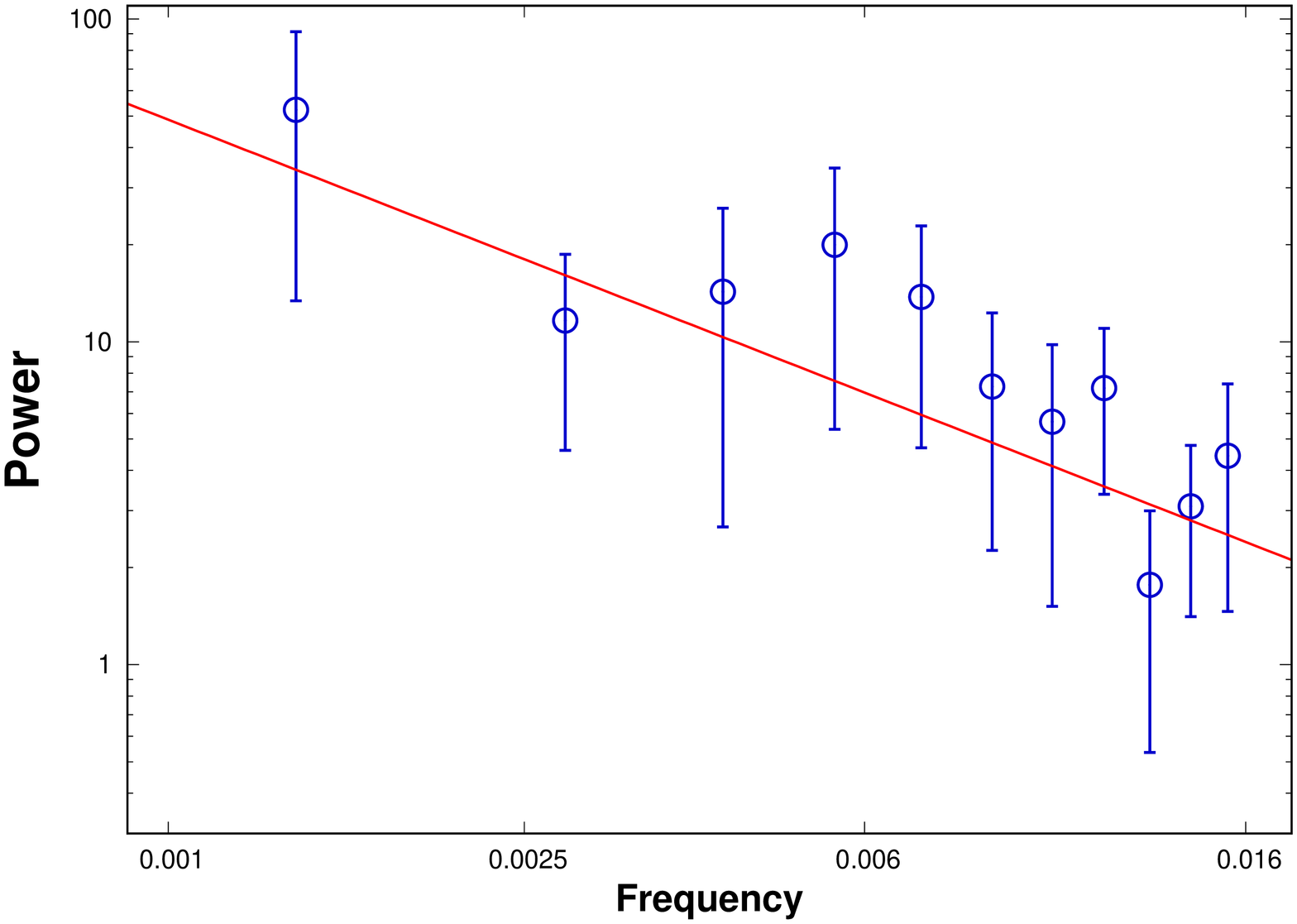}
\caption{Power spectrum of the monthly binned $\gamma$-ray light curve of 3FGL\,J0730.2-1141. The red solid line is the power law fit with $\beta=1.02\pm0.22$.}
\label{fig:psd_fit}
\end{figure}

The  $\gamma$-ray flux distribution for a sample of bright \emph{Fermi}-blazars has been studied in detail using skewness, AD test and histogram fitting  by \cite{Shah2018}. To check the application of our results derived in section \S 3,  we  consider one of the bright \emph{Fermi} blazar viz. 3FGL\,J0730.2-1141 from the sample studied by \cite{Shah2018}. Using the skewness and AD test, \cite{Shah2018} showed that the monthly binned  $\gamma$-ray flux distribution  of this source is log-normal distribution. 
%obtained in more than eight years of observation gives the $\kappa$ and  $TS_{diff}$ of logarithmic of flux distribution  as 0.08 and -0.38 respectively, whereas $\kappa$ and  $TS_{diff}$ of flux distribution it-self are obtained as 1.41 and 2.56 respectively \citep{Shah2018}. These results clearly suggests that the $\gamma$-ray flux  distribution is log-normal. 
The number of flux points (after applying quality cuts) in the considered distribution are 99 with normalised standard deviation, $\sigma_{F,norm}\sim0.68$. The skewness, $\kappa$ and AD TS values of the logarithm of flux distribution are obtained as -0.35 and 0.62 respectively. In order to check the reliability of the AD and skewness test for confirming the log-normality of  flux distribution, we computed the error ratio `R' and power spectra of the monthly binned $\gamma$-ray light curve. The R-value for the observed light curve is obtained as 0.0417. We computed the averaged power spectrum by dividing the monthly binned light curve into four segments, each segment of duration 24 months with a binning time resolution of one month. The power law fit ($P(f)\propto f^{-\beta}$) to the averaged power spectrum, shown in Figure \ref{fig:psd_fit}, results in $\beta=1.02\pm0.22$.  Corresponding to $\beta\sim1.0$ and n=99, the standard deviation on skewness $\sigma_{\kappa}$  and $cv_{5\%}$ are obtained as $\sim$0.25 and $\sim$0.98 respectively (see Table \ref{table:skew_err} and \ref{table:critical_val}). On using the $\kappa$, $\sigma_\kappa$, TS and $cv_{5\%}$, the values of $\Delta \kappa$ and $TS_{diff}$ of the logarithm of flux distribution are obtained as -0.15 and -0.36 respectively, both values suggesting the log-normal behaviour of the flux distribution.  Further, as shown in Figure \ref{fig:skew_frac_error}, for $\beta\sim 1$, $\sigma_{F,norm}\sim0.68$ and $R<0.1$, the available number of flux points, n=99 are sufficient to satisfy the condition required to reject the Gaussianity of flux distribution at 90\% confidence.

\section{Summary and Discussion}\label{sec:conc}

In this work, using simulations, we have studied the efficiency and reliability
of the skewness and AD tests to study whether a  
lightcurve corresponds to a log-normal distribution. The tests are
used to ascertain whether the lightcurve's flux distribution is inconsistent
with a Gaussian one, while the distribution of the log of flux is
consistent with being a normal distribution. Our motivation is to
provide a prescription or guidelines for the effective use of these tests for
this endeavour. The recommended prescription is described below, while
an example of its implementation is given in the previous section.

$\bullet$ For an intrinsically log-normal distribution, the
    efficiency of the skewness and AD tests to reject that the distribution
    is a Gaussian one, is similar. The minimum number of bins required in a
    lightcurve to reject Gaussianity is nearly equal for both the tests.

$\bullet$ It is important to ascertain the nature of the power spectrum of the
lightcurve by estimating the index $\beta$ of the power-law describing the
spectrum. If $\beta \sim 0$ (i.e. white noise) then binning changes the
distribution and since all lightcurves are binned to some extent, estimating
its flux distribution is meaningless. If $\beta \gtrsim 1.5$, then the
variance of the skewness is large and independent of n, and again the skewness and AD tests are not
meaningful.

$\bullet$ If $\beta \sim 0.5$ or $\sim 1$, then we have provided empirical
fits to the standard deviations of the skewness and tabulated critical values for the AD test.
These are different from the ones for white noise and we computed them using
simulations since we could not find these values in the Literature. The
skewness standard deviation and critical values can be used to determine significance by
which the tests reject that the distribution is normal.

$\bullet$ The minimum number of bins required to show that a log-normal distribution is not consistent with a Gaussian one, $N_{90}$ has been computed
as a function of normalised standard deviation and shown in Figures \ref{fig:skew_frac} and  \ref{fig:skew_frac_error} for
different power spectrum index $\beta$ and levels of measurement error.
These can be used {\it a priori} to indicate whether a particular
observation plan will produce lightcurves where these tests can be applied.
Moreover, if a lightcurve having larger number of points shows that its
flux distribution is consistent with a Gaussian one, then one can say with
some confidence that its distribution is not a log-normal one.

$\bullet$ The presence of Gaussian measurement errors increases $N_{90}$ and
we recommend that the error ratio, R  which is the ratio of
the standard deviation of error to the intrinsic standard deviation of flux, to be $ < 0.2$. We suggest that
the data should be binned in time till $R < 0.2$. One needs to take care while constructing a histogram for characterising a PDF shape. The histogram bin width should be chosen such that the width of bin is larger than the measurement error in that bin. Smaller histogram bin-width will introduce bias in the shape of PDF, as the flux point with large error can not be located in single bin. Larger bin-width in the histogram will be required for the larger measurement errors, which some times produces histogram bins with empty counts. Therefore, in order to reduce the measurement error, one should bin the data till the error ratio, $R$ is optimised.  Our work suggest $R< 0.2$.

$\bullet$ For a system having an intrinsic log-normal distribution, the skewness and AD test will accept the normality of the logarithm of the flux at more than 95\% confidence level. However if there is significant Gaussian measurement error in the flux, the skewness and AD test may not accept the normality of the logarithm of the flux at 95\% confidence level. We show that for an error fraction of $R\lesssim0.2$, this would require more than $10^5$ data points. Thus for most practical considerations this effect would be negligible.

$\bullet$ The average skewness of the flux of a log normal distribution
differs from its theoretical value for finite number of bins. Moreover, its
variance is large. Hence, We do not recommend that the value of the skewness
of the flux to be used to show that the distribution is a log-normal one.

This analysis is pertinent to the case when one has a lightcurve with
moderate number of bins say of the order of hundreds. For a larger data set,
one can directly fit the histogram of the flux (or the log of the flux) with
different distributions to ascertain its nature. Naturally, for a data
set with moderate number of bins, the power spectra estimate will not be
precise and the power spectra index $\beta$ would only be known
approximately. That is the reason why we have only considered
$\beta \sim 0.5$ 
and $1.0$, and not undertaken numerically expensive analysis for finer values
of $\beta$. This implies that any result obtained from the skewness and AD
tests should come with the caveat that they depend on the uncertainty of
estimating the power spectrum. For similar reasons, we have not considered
the more generic situation when the power spectrum can be described as a
broken power-law. We note that the variance of the skewness and the
AD critical values listed in this work are for power-law power spectra
and are not applicable for a broken power-law. For typical cases, where
a flat power spectrum ( e.g. $\beta \sim 0.5$) breaks into a
steeper one (e.g. $\beta \sim 1.5$) beyond a break frequency, we recommend
that the lightcurve be rebinned to a bin size corresponding to the
break frequency and then apply the skewness and AD tests. Since the
power-law index is typically steep beyond the break frequency, rebinning will
not lead to significant loss of information and in general will decrease
the measurement noise level.

As an important aside, we have addressed a conceptually paradoxical
issue regarding non-Gaussian flux distributions.  While Gaussian distributions
remain Gaussian on addition, i.e. Gaussian flux distribution remain Gaussian
when the lightcurve is rebinned, this is in general not true for other
 distributions such as a log-normal one. Thus, it is not clear
why several astrophysical systems display log-normal distributions when
the lightcurves they are estimated from, are always binned to some
time-bin? We resolve this issue by showing that if the power spectra
of the lightcurves can be described by a power law with index $\beta \gtrsim 0.5$,
then the nature of the distribution is invariant to rebinning. Thus for
a system to be scale free in time and have a log-normal flux distribution its 
power spectra must be steeper than 0.5. This insight may have important
consequences for the model development and understanding of systems with
log-normal distributions.

Understanding the nature of flux distribution using first analysis tools
like skewness and AD that can be applied to moderate size data sets, is an
important step laying the foundation for more elaborate studies. The results
presented here provide a prescription and guidelines for the effective
and reliable use.

\section{ACKNOWLEDGEMENTS}
We thank the anonymous referee for his critical assessment of our work. The referee comments and suggestions have helped us in improving this manuscript. We thank a UGC-UKIERI Thematic Partnership for support. We acknowledge the use of Fermi-LAT
data provided by Fermi Science Support Center (FSSC).

\section{Data availability}
The Python codes and data used in this article will be shared on reasonable request to the corresponding author, Zahir Shah (email: zahir@iucaa.in or shahzahir4@gmail.com).

\bibliographystyle{mnras}
\bibliography{references}

\appendix
\section{Skewness}\label{apx:SS}
\begin{figure}
\centering
\includegraphics[width=0.34\textwidth,angle=270]{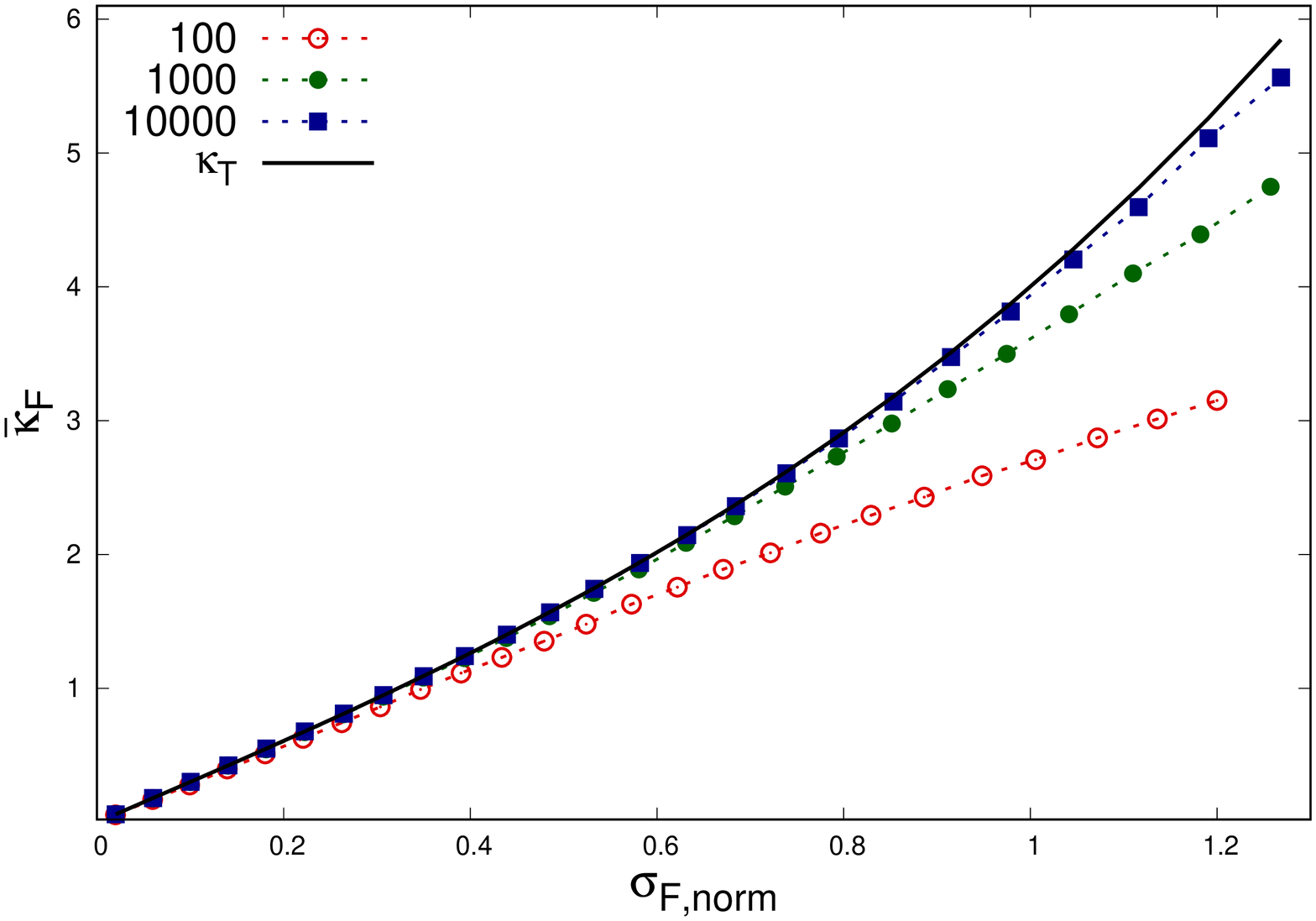}
\caption{Average of the skewness of flux distribution ($\bar{\kappa}_F$) against normalised standard deviation of flux distribution $\sigma_{F,norm}$. The  solid curve with  black colour corresponds to the theoretical curve. The dotted curves are for different values of length of light curve, n.}
\label{fig:skew_vs_sigma}
\end{figure}

\begin{figure}
\centering
\includegraphics[width=0.34\textwidth,angle=270]{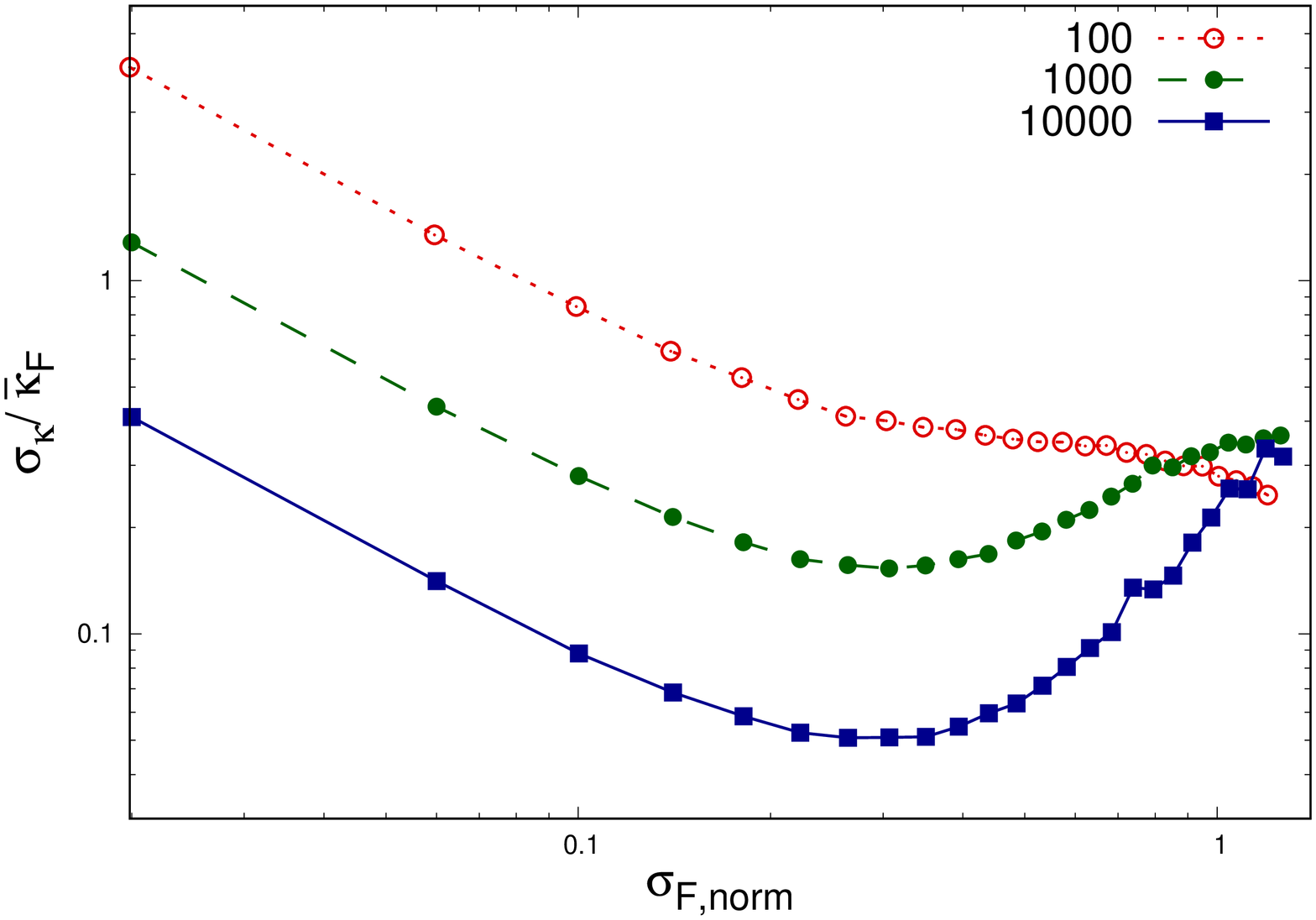}
\caption{ The standard deviation of the skewness divided by the mean of skewness ($\sigma_\kappa/\bar{\kappa_F}$) vs the normalised standard deviation of flux distribution $\sigma_{F,norm}$ for n=100, 1000, 10000}
\label{fig:std_skew_var}
\end{figure}

A possible way to test the log-normality of a flux distribution is
to compare the computed skewness of the distribution with the
theoretical value given by Equation \ref{eq:pred_skew}. However, we find that for
a moderate number of data points of hundreds, this is not a reliable method.
We  simulated  light curves corresponding to a log-normal
flux distribution for different sets of $\sigma_{LF}$ and $n$ (see
section \ref{sec:skew_AD_white} for details), 10000 light curves for
each set of $\sigma_{LF}$ and $n$. We measured the skewness value
$\kappa_F$ for all such light curve, and calculated their average,
$\bar{\kappa}_F$. Figure
(\ref{fig:skew_vs_sigma}) shows  $\bar{\kappa}_F$
as a function of  $\sigma_{F,norm}$ 
using dotted lines with  symbols representing different lengths of the lightcurve, $n$. The solid curve
is the theoretical skewness calculated using Equation
(\ref{eq:pred_skew}). The Figure clearly shows that $\bar{\kappa}_F$ is
significantly smaller than the theoretical values for small $n$ and this
deviation increases with the standard deviation of the lightcurve. Only
asymptotically with large $n$ does $\bar{\kappa}_F$ match with the
theoretical values. Thus, comparison of the computed skewness with
the theoretical value cannot be done, for moderate length lightcurves.
While it is possible to empirically fit the expected $\bar{\kappa}_F$
as a function of $n$ and normalised flux standard deviation, $\sigma_{F,norm}$ ,
this may not be warranted because of the large variation of $\kappa_F$. This
is shown in Figure (\ref{fig:std_skew_var}), where the standard deviation
of the skewness (normalised by the average) is shown as function of
$\sigma_{F,norm}$ for different $n$. For $n < 1000$, the variation
of $\kappa_F$ is always more than 20\%, which will not allow for any
meaningful inference.
  \\

\end{document}